\documentclass[%
superscriptaddress,
preprint,
 amsmath,amssymb,
 prd,
]{revtex4-1}

\usepackage{graphicx}
\usepackage{dcolumn}
\usepackage{bm}

\begin{document}

\preprint{APS/123-QED}

\title{Theoretical Calculation of coherent Laue-case conversion between X rays and ALPs\\ for an X-ray LSW experiment}

\author{T.~Yamaji}
 \email{yamaji@icepp.s.u-tokyo.ac.jp}
\affiliation{%
Department of Physics, Graduate School of Science, The University of Tokyo, 7-3-1 Hongo, Bunkyo, Tokyo 113-0033, Japan\\
}%
\author{T.~Yamazaki}
\affiliation{
High energy Accelerator Research Organization, KEK, 203-1 Tokai-mura, Naka-gun, Ibaraki 319-1106, Japan\\
}%
\author{K.~Tamasaku}
\affiliation{%
RIKEN SPring-8 Center, 1-1-1 Kouto, Sayo-cho, Sayo-gun, Hyogo 679-5148, Japan\\
}%
\author{T.~Namba}
\affiliation{%
International Center for Elementary Particle Physics, The University of Tokyo, 7-3-1 Hongo, Bunkyo, Tokyo 113-0033, Japan\\
}%

\date{\today}

\begin{abstract}
Single crystals have high atomic electric fields as much as $10^{11}$ V/m, which correspond to magnetic fields of $\sim 10^3$ T. These fields can be utilized to convert X rays into Axion Like Particles (ALPs) coherently similar to X-ray diffraction. In this paper, we perform the first theoretical calculation of the Laue-case conversion in crystals based on the Darwin dynamical theory of X-ray diffraction. The calculation shows that the Laue-case conversion has longer interaction length than the Bragg case, and that  ALPs in the keV range can be resonantly converted by tuning an incident angle of X rays. ALPs with mass up to $\mathcal{O}(10\ {\rm keV})$ can be searched by Light-Shining-through-a-Wall  (LSW) experiments at synchrotronX-ray facilities.
\end{abstract}

\maketitle

\section{Introduction}
Some theories beyond the Standard Model predict additional particles which have weak couplings to photons. 
The unknown particles of this sort include Axion Like Particles (ALPs) \cite{bib:ALPs1,bib:ALPs2,bib:ALPs3}, which are particles with similar properties to axion \cite{bib:PQ1,bib:PQ2,bib:WW1,bib:WW2}. The standard axion is a Nambu-Goldstone boson associated with an additional $U(1)$ symmetry, which is firstly motivated to provide a solution for the strong-$CP$ problem  \cite{bib:PQ1,bib:PQ2}. In addition to the theoretical point of view, ALPs are of astronomical interest. ALPs are one of viable dark matter candidates \cite{bib:DM1,bib:DM2}, and they can provide possible explanation for astronomical observations such as the $\gamma$-ray transparency of the Universe \cite{bib:trans1,bib:trans2} and the stellar evolution \cite{bib:stellar1,bib:stellar2}. In particular, ALPs in the keV range can be related to the anomalies of solar activities such as coronal heating \cite{bib:keV_motivation} and an unidentified X-ray emission line around 3.5 keV recently detected in the galaxy clusters \cite{bib:galaxy}. Search for ALPs in the keV range has a particular importance in astronomy as well as elementary particle physics.

ALPs and photons can transform into each other via a mixing process (Primakoff effect \cite{bib:primakov1}), which may be described by the Lagrangian density, 
\begin{eqnarray}
\mathcal{L}_{\rm int}=-\frac{g_{a\gamma\gamma}}{4}\mathcal{F}_{\mu\nu}\widetilde{\mathcal{F}}^{\mu\nu}a=g_{a\gamma\gamma}{\bf E}\cdot{\bf B}a,
\end{eqnarray}
where $\mathcal{F}_{\mu\nu}$ is the electromagnetic field strength, $\widetilde{\mathcal{F}}^{\mu\nu}=\frac{1}{2}\epsilon^{\mu\nu\rho\sigma}\mathcal{F}_{\rho\sigma}$ is its dual, $a$ represents the ALPs fields, $g_{a\gamma\gamma}$ is the coupling constant, and ${\bf E}\cdot{\bf B}$ is the odd-parity product of electromagnetic fields. The standard axion has proportionality between its mass and coupling constant \cite{bib:KSVZ1,bib:KSVZ2,bib:DFSZ1,bib:DFSZ2}. On the other hand, the ALPs' mass $m_a$ and $g_{a\gamma\gamma}$ are considered to be not bound to each other.

Although upper limits on $g_{a\gamma\gamma}$ in a broad mass range are obtained by solar axion searches, their limits inevitably depend on a solar model \cite{bib:ssm1,bib:ssm2,bib:ssm3,bib:ssm4,bib:ssm5,bib:ssm6} and its magnetic activity \cite{bib:magnetic}. For example, the limit can be relaxed by postulating that the ALPs' mass and coupling depend on the environmental condition such as the temperature and the matter density \cite{bib:ssm5}. Pure-laboratorial experiments have a big advantage of model-independent searches for ALPs. Various laboratorial experiments based on the Primakoff effect have been carried out by using the LSW scheme \cite{bib:LSW1}. 
The LSW scheme converts photons (${\bf E}$) into ALPs by mixing them in the presence of an external magnetic field (${\bf B}$) of $\sim \mathcal{O}(1)$ T. The generated ALPs pass through a shielding wall that blocks unconverted photons. Some of the ALPs are subsequently re-converted into real photons via an inverse process in another magnetic field. The mixing takes place when ALPs' mass $m_a$ is much less than the energy of real photons. Most LSW experiments are performed by using optical lasers as photon sources \cite{bib:LSWo1,bib:LSWo2,bib:LSWo3,bib:LSWo4,bib:LSWo5,bib:LSWo6,bib:LSWo7,bib:LSWo8,bib:LSWo9,bib:LSWo10,bib:LSWo11}, and experiments using X-ray sources \cite{bib:LSWXE,bib:inada} are recently performed to probe heavier ALPs. The sensitive ALPs' mass of these X-ray experiments is up to $\mathcal{O}$(1~eV). 

\begin{figure}
  \centering
  \includegraphics[width=0.5\textwidth, angle=0]{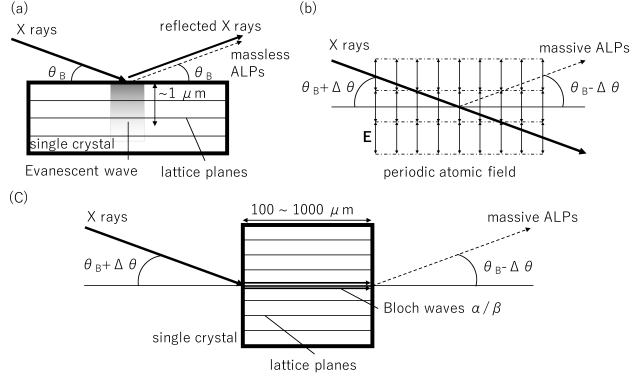}
  \caption{Schematics of an X ray-ALP conversion by atomic electric fields. (a) The Bragg-case conversion at the Bragg angle $\theta_{\rm B}$ \cite{bib:braggtheory}. The X-ray penetration length is $\sim 1\ {\rm \mu m}$. (b) The conversion away from the Bragg condition studied in Ref.~\cite{bib:simpletheory}. The study takes into account ALPs' mass and the detuning angle $\Delta\theta$. Massive ALPs can be generated only when the incident angle is detuned from $\theta_{\rm B}$. The crystal is approximated just as a periodic field in the calculation. (c) The Laue-case conversion which we consider in this paper. The effects of X-ray diffraction and ALPs' mass are considered at the same time.}
   \label{fig:conv_types}
\end{figure}

It is well-known that there are extremely high electric fields in crystals. The fields are as high as $10^{11}$~V/m, which correspond to magnetic fields of $\sim 10^3$ T. These electric fields can also be used to transform X rays into ALPs or vice versa. The feasibility to use the electric fields in crystals for the conversion has been studied in Ref.~\cite{bib:braggtheory} and \cite{bib:simpletheory}. These studies propose coherent X ray-ALP conversion similar to X-ray diffraction as shown in  Fig.~\ref{fig:conv_types}. The conversion in Bragg-case reflection (Fig.~\ref{fig:conv_types}~(a)) is firstly considered in the framework of the Darwin dynamical theory of X-ray diffraction by {\it Buchm\"{u}ller} and {\it Hoogeveen}  \cite{bib:braggtheory}. The Bragg-case reflection takes place when X rays fall on a crystal with reflecting lattice planes parallel to its surface (Bragg-case). The reflection converts some X rays into ALPs at the Bragg angle $\theta_{\rm B}$. The calculated conversion probability for nearly massless ALPs is as follows,
\begin{equation}
P_{a\leftrightarrow\gamma}=\left(\frac{1}{2}g_{a\gamma\gamma}E_{\rm T}L_{\rm B}{\rm cos}\theta_{\rm B}\right)^2,\label{eq:bragg_prob}
\end{equation}
where $E_{\rm T}$ is an effective electric field shown afterwards, and $L_{\rm B}$ is an X-ray penetration length under the Bragg condition. 
Solar axion searches using the crystal fields have been proposed \cite{bib:solarAP1,bib:solarAP2} and performed \cite{bib:solarAE1,bib:solarAE2,bib:solarAE3,bib:solarAE4,bib:solarAE5}. These experiments use a crystal detector itself as a converter from solar axions into X rays. However, LSW experiments using atomic electric fields have not been performed because Bragg-case diffraction in crystals reduces the X-ray penetration length $L_{\rm B}\sim 1\ {\rm \mu m}$ and the production efficiency of ALPs significantly. It was also suggested in Ref.~\cite{bib:braggtheory} that the conversion is more effective when reflecting lattice planes are perpendicular to crystal surfaces (Laue-case). Later, {\it Liao} takes into account nonzero ALPs' mass in Ref.~\cite{bib:simpletheory} by approximating crystals as periodic electric field (Fig.~\ref{fig:conv_types}(b)). The study showed that ALPs in the keV range can be continuously searched by scanning incident angles of X rays. However, some effects of X-ray absorption and scattering which exist in real crystals are ignored in the previous calculation. Rigorous calculation including these effects is required to consider realistic experiments using atomic electric fields. 

In this paper, we propose a new conversion geometry with reflecting lattice planes perpendicular to the crystal surfaces (Laue-case) as shown in Fig.~\ref{fig:conv_types}(c), and calculate the conversion efficiency. This calculation is performed in consideration of X-ray diffraction, X-ray absorption and nonzero $m_a$. 

\section{Laue-case conversion between X rays and ALPs in crystals}
In this section, we perform a theoretical calculation of the Laue-case conversion between X rays and ALPs in a crystal. The calculation is based on the Darwin dynamical theory \cite{bib:book}.

We set up recurrence formulae for amplitudes of X rays and ALPs at first. In the next place, we provide a solution of them by using the Born approximation. Finally, we examine the effect of X-ray diffraction on the X ray-ALP conversion. In this paper, we apply natural units with Lorentz-Heaviside units.

\subsection{Recurrence relation between amplitudes of X-ray and ALP waves}
We consider Laue-case conversion in a crystal between X rays with a momentum of ${\bf k}_\gamma$ and massive ALPs with a momentum of ${\bf k}_a$. 
\begin{figure}
  \centering
  \includegraphics[width=0.5\textwidth, angle=0]{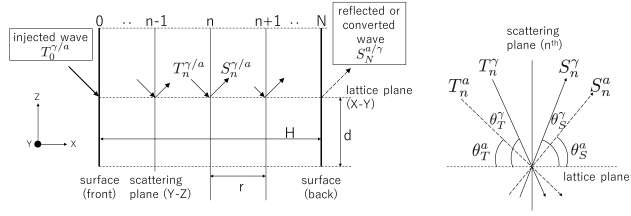}
  \caption{Schematics of the X ray-ALP conversion in the framework of the Darwin dynamical theory. Left: The crystal geometry for the calculation. Atoms in the single crystal with a thickness of $H$ are divided into scattering planes parallel to the crystal surface. Scattering planes have an arbitral spacing of $r$. The overall conversion probability can be calculated by solving recurrence relations between transmitted and reflected amplitudes $T_n^{\gamma,a},\ S_n^{\gamma,a}$ at each scattering planes. Right: The definition of $\theta^{\gamma/a}_{T/S}$}
   \label{pic:darwin}
\end{figure}
For the convenience of the calculation, we consider the geometry shown in Fig.~\ref{pic:darwin}, where X-ray diffraction and the conversion take place in the X-Z plane, the crystal surfaces are parallel to the Y-Z plane, and the reflecting lattice planes with the spacing of $d$ are parallel to the X-Y plane. Since the conversion is the most efficient for $\sigma$-polarization, we postulate that X rays are linearly polarized in Y direction. We divide all atoms in the crystal into virtual scattering planes parallel to the crystal surfaces. Their spacing $r$ and number $N$ are dummy variables subject to the constraint $rN=H$, where $H$ is the thickness of the crystal. These dummy variables vanish at the final phase of the calculation. 

The Laue-case conversion originates from a transition between four waves in the crystal, transmitted/reflected X rays ($T^\gamma/S^\gamma$) and ALPs ($T^a/S^a$). The transition takes place due to X-ray scattering and X ray-ALP conversion by atoms on the scattering planes. The total conversion efficiency can be calculated by solving recurrence relations of transmitted/reflected amplitudes $T^\gamma_n/S^\gamma_n$ and $T^a_n/S^a_n$ at $n$-th scattering planes. 
The conversion and scattering take place coherently when angles between lattice planes and the four waves satisfy the Fresnel diffraction condition as follows \cite{bib:book},
\begin{eqnarray}
k_\gamma{\rm sin}\theta^\gamma_T+k_\gamma{\rm sin}\theta^\gamma_S&=&k_\gamma{\rm sin}\theta^\gamma_T+k_a{\rm sin}\theta^a_S\nonumber \\
&=&k_a{\rm sin}\theta^a_T+k_\gamma{\rm sin}\theta^\gamma_S=q_T,
\label{eq:fresnel}
\end{eqnarray}
where $\theta^{\gamma/a}_{T/S}$ are the angles of transmitted/reflected X rays/ALPs,  $k_{\gamma/a}\equiv|{\bf k}_{\gamma/a}|$ are the momentum of X rays/ALPs, and $q_{\rm T}\equiv\frac{2\pi}{d}$ is the reciprocal lattice spacing. The Fresnel condition determines the angles $\theta^{\gamma/a}_{T/S}$ and the momentum transfer ${\bf q}$ as a function of the incident angle $\theta^{\gamma/a}_T$. When $\theta^\gamma_T=\theta^\gamma_S$, the condition is reduced to the Bragg condition $2k_\gamma{\rm sin}\theta_{\rm B}=q_T$. When the injection angle of X rays is detuned from $\theta_{\rm B}$ by $\Delta\theta$, $\theta^\gamma_T=\theta_{\rm B}+\Delta\theta$, $\theta^{\gamma/a}_S\simeq\theta_{\rm B}-\Delta\theta$ and $\theta^{a}_T\simeq\theta_{\rm B}+\Delta\theta$. Since the left hand side of Eq.~(\ref{eq:fresnel}) corresponds to the momentum transfer in the $z$ direction, ${\bf q}\cdot \hat{\bf z}=q_T$ where $\hat{\bf z}$ is the unit vector in the $z$ direction. The momentum transfer in the $x$ direction is calculated from Eq.~(\ref{eq:fresnel}) as
\begin{equation}
({\bf q}\cdot \hat{\bf x})_{ij}=\begin{cases}
\frac{2q_T(k_\gamma{\rm sin}\theta^\gamma_T-\frac{q_T}{2})}{2k_\gamma {\rm cos}\theta^\gamma_T} & (T^\gamma\ {\rm to}\  S^\gamma,\ \gamma\gamma) \\
-\frac{m_a^2-2q_T(k_\gamma{\rm sin}\theta^\gamma_T-\frac{q_T}{2})}{2k_\gamma {\rm cos}\theta^\gamma_T} & (T^\gamma\ {\rm to}\ S^a,\ {\rm \gamma a}) 
\end{cases},
\label{eq:qx}
\end{equation}
where the subscript $(ij)$ is $\gamma\gamma$ or $\gamma a$, $\hat{\bf x}$ is the unit vector in the $x$ direction, and the approximation $k_\gamma{\rm cos}\theta^\gamma_T\simeq k_\gamma{\rm cos}\theta^\gamma_S\simeq k_a{\rm cos}\theta^a_S$ is used in the denominator. Since ${\bf q}\cdot \hat{\bf x}$ is much smaller than ${\bf q}\cdot \hat{\bf z}$ when $\Delta\theta\sim \mathcal{O}(1\ {\rm mrad})$, we approximate ${\bf q}\simeq q_T\hat{\bf z}$. The momentum transfer in the x direction is directly related to phase differences between the four waves as will be shown later.

The four waves which satisfy the Fresnel condition are mixed with each other by X-ray scattering and X ray-ALP conversion on the scattering planes. For example, the amplitude of transmitted X-ray wave, $T^\gamma_{n+1}$, is made up of the following three components: transmission and forward scattering of $T^\gamma_{n}$, scattering of $S^\gamma_{n-1}$ and conversion from $S^a_{n-1}$. On the contrary, the amplitude of reflected X-ray, $S^\gamma_{n}$, is made up of reflection of $T^\gamma_{n}$, transmission and forward scattering of $S^\gamma_{n-1}$ and conversion from $T^a_{n}$. The amplitudes of ALPs can be described in the same way, except that ALPs are not scattered by scattering planes. These recurrence relations between $T^{a,\gamma}_n$ and $S^{a,\gamma}_n$ are shown as,
\begin{eqnarray}
\label{eq:recurrence1}
T^\gamma_{n+1}&=&T^\gamma_n(1+i\eta_{T0})e^{-i\phi^\gamma_T}+S^\gamma_{n-1}(i\eta_T)e^{-i(\phi^\gamma_S+\phi^\gamma_T)} \nonumber\\
&&+S^a_{n-1}(i\zeta^\prime_{ST})e^{-i(\phi^a_S+\phi^\gamma_T)},\\
\label{eq:recurrence2}
S^\gamma_n&=&T^\gamma_n(i\eta_S)+S^\gamma_{n-1}(1+i\eta_{S0})e^{-i\phi^\gamma_S}+T^a_n(i\zeta^\prime_{TS}),\\
\label{eq:recurrence3}
T^a_{n+1}&=&T^a_ne^{-i\phi^a_T}+S^\gamma_{n-1}(i\zeta_{ST})e^{-i(\phi^\gamma_S+\phi^a_T)},\\
\label{eq:recurrence4}
S^a_n&=&S^a_{n-1}e^{-i\phi^a_S}+T^\gamma_n(i\zeta_{TS}).
\end{eqnarray}
The variables in the formulae are defined as follows \cite{bib:braggtheory},
\begin{eqnarray} 
\eta_i&\equiv& \eta_i({\bf q})=\frac{r_{e}M\lambda_\gamma f_\gamma({\bf q})}{{\rm cos}\theta^\gamma_i},\\
\eta_{i0} &\equiv& \eta_i({\bf 0}),\\
\zeta_{jk}&\equiv&\frac{g_{a\gamma\gamma}M\lambda_a f_a({\bf q})}{4\pi {\rm cos}\theta^a_k}{\rm sin}(\theta^\gamma_j+\theta^a_k),\\
\zeta^\prime_{jk}&\equiv&\frac{g_{a\gamma\gamma}M\lambda_\gamma f_a({\bf q})}{4\pi {\rm cos}\theta^\gamma_k}{\rm sin}(\theta^a_j+\theta^\gamma_k),\\
\phi^l_{i}&\equiv&k_l r {\rm cos}\theta^l_i,
\end{eqnarray}
where the subscripts are $i=T/S$, $l=\gamma/a$, $(j, k)$=(T,~S)/(S,~T), $r_{\rm e}$ is the classical electron radius, $M=\frac{F_c}{V}r$ is the effective number density of atoms on the scattering plane, $F_c$ is the coefficient of the crystal structure factor, $V$ is the volume of the unit lattice cell, $\eta_i$ are the X-ray scattering amplitudes, $\eta_{i0}$ are the X-ray forward scattering amplitudes, $\zeta_{jk}/\zeta^\prime_{jk}$ are the X ray-ALP/ALP-X ray conversion amplitudes, $\phi^l_{i}$ are the phase changes of four waves during the propagation among scattering planes, $\lambda_{\gamma,a}$ are the wavelengths of X rays and ALPs. $f_{\gamma,a}({\bf q})$ are the atomic form factor and a conversion form factor analogous to it, 
\begin{eqnarray}
f_\gamma({\bf q})&=&\frac{1}{e}\int d^3{\bf r}\rho({\bf r}){\rm exp}(i{\bf q}\cdot {\bf r}),\\
f_a({\bf q})&=&k_{\rm a}^2\int d^3{\bf r}\phi({\bf r}){\rm exp}(i{\bf q}\cdot {\bf r}),
\end{eqnarray}
where $\rho ({\bf r}), \phi({\bf r})$ are the electron charge density and the electric potential of an atom, $e$ is the elementary charge. The forward form factor $f_a({\bf 0})$ vanishes since these form factors have the relation $f_a({\bf q})\propto[Z-f_\gamma({\bf q})]$ and $f_\gamma({\bf 0})=Z$, where $Z$ is the atomic number of the crystal.

\vspace{\baselineskip}

\subsection{Solution of the recurrence formulae}
Now we deduce the conversion amplitude by solving the recurrence relations, Eqs.~(\ref{eq:recurrence1}-\ref{eq:recurrence4}). We consider the conversion from X rays to ALPs under the boundary condition of  $T^\gamma_0=T_0$ and $T^a_0=0$ on the incident surface. The re-conversion contribution from ALPs in Eqs.~(\ref{eq:recurrence1}, \ref{eq:recurrence2}) are $\mathcal{O}(\zeta^2)$ since $T^a_n, S^a_n$ are estimated to be $\mathcal{O}(\zeta)$ by Eqs.~(\ref{eq:recurrence3}, \ref{eq:recurrence4}). We ignore the re-conversion contribution, which corresponds to the Born approximation. By the approximation, the recurrence relations between $T^\gamma_n$ and $S^\gamma_n$ become the same as the conventional Laue-case X-ray diffraction. Their solution can be deduced by using characteristic functions as follows \cite{bib:book},
\begin{eqnarray}
\label{eq:solution}
T^\gamma_n&=&T_0(C_\alpha x_\alpha^n+C_\beta x_\beta^n).
\end{eqnarray}
The parameters in Eq.~(\ref{eq:solution}) are defined as 
\begin{eqnarray}
C_{\alpha/\beta}&\equiv&\frac{1}{2}\left[1\pm \frac{{\rm sin}\Delta \phi_\gamma (1+i\eta_0)}{u}\right],\label{eq:c}\\
x_{\alpha/\beta}&\equiv&e^{-i\phi_\gamma}[(1+i\eta_0){\rm cos}\Delta \phi_\gamma\mp iu],\label{eq:x}\\
u&\equiv&[\eta^2+(1+i\eta_0)^2{\rm sin}^2\Delta \phi_\gamma]^\frac{1}{2},\label{eq:u}\\
\eta&\equiv&\eta({\bf q})= \frac{r_e M\lambda_\gamma f_\gamma({\bf q})}{{\rm cos}\theta_{\rm B}} \label{eq:eta},\\
\eta_0&\equiv& \eta({\bf 0})=\frac{r_e M\lambda_\gamma Z}{{\rm cos}\theta_{\rm B}} \label{eq:eta0},\\
\phi_\gamma&\equiv&\frac{1}{2}(\phi^\gamma_T+\phi^\gamma_S),\\
\Delta\phi_\gamma&\equiv&\frac{1}{2}(\phi^\gamma_T-\phi^\gamma_S)=-\frac{r}{2}({\bf q}\cdot \hat{\bf x})_{\gamma\gamma},\label{eq:dphi_gg}
\end{eqnarray}
where the approximation $\sqrt{\eta_{T}\eta_{S}}\simeq \eta$, $\sqrt{\eta_{T0}\eta_{S0}}\simeq \eta_0$ and $\eta_{T0}+\eta_{S0}\sim 2\eta_0$ are used. $\Delta\phi_\gamma$ is related to the momentum transfer in the $x$ direction shown by Eq.~(\ref{eq:qx}), which represents the deviation from the Bragg condition. The indexes $\alpha/\beta$ label X-ray standing waves (Bloch wave $\alpha/\beta$) explained later. The corresponding representation for the ALP phases is introduced as
\begin{eqnarray}
\phi_a&\equiv&\frac{1}{2}(\phi^\gamma_T+\phi^a_S),\\
\Delta\phi_a&\equiv&\frac{1}{2}(\phi^\gamma_T-\phi^a_S)=-\frac{r}{2}({\bf q}\cdot \hat{\bf x})_{\gamma a}.
\end{eqnarray}
The phase difference $\Delta\phi_a$, which is related to a resonant condition for the X ray-ALP conversion, has the following relation,
\begin{equation}
\Delta\phi_a N=\frac{L}{4 k_\gamma}\left[m_a^2-2q_T\left(k_\gamma{\rm sin}\theta^\gamma_T-\frac{q_T}{2}\right)\right],
\label{eq:delta_phi_a}
\end{equation}
where we use Eq.~(\ref{eq:qx}), and $L=\frac{H}{{\rm cos}\theta^\gamma_T}$ is the X-ray path length in the crystal. It is important to note that the resonance of X ray-ALP conversion does not necessarily require the Bragg condition as required in Ref.~\cite{bib:braggtheory}. From these parameters and Eqs.~(\ref{eq:recurrence4}, \ref{eq:solution}), we can obtain the amplitude of reflected ALPs at the $N$-th scattering plane as,
\begin{eqnarray}
S^a_N&=&(i\zeta_{TS})\sum_{n=0}^NT^\gamma_ne^{-i(N-n)\phi^a_S}\nonumber\\
&=&T_0(i\zeta_{TS})e^{-iN\phi^a_S}\sum_{j=\alpha/\beta}C_j\frac{1-(x_je^{i\phi_S^a})^{N+1}}{1-x_je^{i\phi_S^a}}.
\label{eq:as}
\end{eqnarray}
The parameters $\Delta\phi_{\gamma/a}, \eta_{(0)}$ are proportional to the spacing of scattering planes, $r$. Since $r$ and $N$ are dummy variables, higher order contributions of $r$ can be ignored by taking $r\rightarrow 0$ with $rN=H$. When we ignore terms higher than $\mathcal{O}(r^2)$, Eqs.~(\ref{eq:c}, \ref{eq:x}, \ref{eq:u}) may be simplified as
\begin{eqnarray}
\label{eq:c_approx}
C_{\alpha/\beta}&=&\frac{1}{2}\left(1\pm \frac{\Delta\phi_\gamma}{u}\right),\\
x_{\alpha/\beta}e^{i\phi^a_S}&=&1+i(\eta_0\mp u+\Delta\phi_\gamma-2\Delta\phi_a),\label{eq:x_approx}\\
\label{eq:u_approx}
u&=&[\eta^2+\Delta\phi_\gamma^2]^{\frac{1}{2}}.
\end{eqnarray}
By using Eqs.~(\ref{eq:c_approx}, \ref{eq:x_approx}, \ref{eq:u_approx}) and the approximation $\lim_{r\to 0,\ rN={\rm const}} (1+r)^N={\rm e}^{rN}$, Eq.~(\ref{eq:as}) can be simplified to be

\begin{widetext}
\begin{eqnarray}
S^a_N=-\frac{T_0}{2}(\zeta_{TS} N)e^{-iN\phi^a_S}\sum_{\pm}\left[\left(1\pm\frac{\Delta\phi_\gamma}{u}\right)\frac{1-{\rm exp}(i(\eta_0\mp u+\Delta \phi_\gamma-2\Delta \phi_a)N)}{(\eta_0\mp u+\Delta \phi_\gamma-2\Delta \phi_a)N}\right].
\label{eq:xtoalps}
\end{eqnarray}
\end{widetext}
From the above equation, the conversion probability can be calculated as $P_{\gamma\to a}=\left|\frac{S^a_N}{T_0}\right|^2$. The re-conversion probability from ALPs to X rays takes the same form as it, except that $\zeta_{TS}$ is replaced by $\zeta^\prime_{TS}$.

In the following subsections, we consider the conversion probability in the two cases, where the Bragg condition is fulfilled and where the incident angle is away from the Bragg angle.

\newcommand{\ltsim}{\protect\raisebox{-0.5ex}{$\:\stackrel{\textstyle <}{\sim}\:$}} 
\subsubsection{Under the Bragg condition}
The most peculiar effect of Laue-case conversion takes place under the Bragg condition ($\theta^\gamma_T=\theta_{\rm B}$ and $\Delta\phi_\gamma=0$). The conversion amplitude in this case is composed of two components with the attenuation lengths of $L_{\rm att, \alpha/\beta}=\frac{L_{\rm att}}{1\mp \kappa}$, where $L_{\rm att}$ is the normal X-ray attenuation length of the crystal and $\kappa\equiv\frac{{\rm Im}f_\gamma({\bf q})}{{\rm Im}f_\gamma({\bf 0})}$. 
The components originate from standing waves called Bloch waves $\alpha$ (the first term of Eq.~(\ref{eq:xtoalps})) and $\beta$ (the second term) due to X-ray diffraction. Since $\kappa$ is close to unity, the Bloch wave $\alpha$ has much longer attenuation length than $L_{\rm att}$, which is called as Borrmann effect \cite{bib:borrmann1,bib:borrmann2}. The amplitudes of the Bloch waves $\alpha/\beta$ in Eq.~(\ref{eq:xtoalps}) has their maxima under the following condition,
\begin{eqnarray}
({\rm Re}(\eta_0\mp \eta)-2\Delta\phi_a)N=0.
\end{eqnarray}
The condition is equivalent to the resonant ALPs' mass of 
\begin{eqnarray}
m_a^2 = m_\gamma^2 \mp \Delta m_\gamma^2,\label{eq:ma_bragg},
\label{eq:resonant_bragg}
\end{eqnarray}
where we use Eqs.~(\ref{eq:eta}, \ref{eq:eta0}, \ref{eq:delta_phi_a}), $m_\gamma\equiv\sqrt{\frac{4\pi r_e F_c Z}{V}}$ is the plasma frequency of the crystal, and $\Delta m_\gamma\equiv m_\gamma\sqrt{\frac{{\rm Re}f_\gamma({\bf q})}{{\rm Re}f_\gamma({\bf 0})}}$ is its modification caused by X-ray scattering. From the expression Eq.~(\ref{eq:xtoalps}), these resonances have a range of resonant ALPs' mass (a full width at half maximum of $(P_{\gamma\leftrightarrow a})^2$) which can be described as 
\begin{eqnarray}
&&\left|m_a^2-(m_\gamma^2 \mp \Delta m_\gamma^2)\right|\nonumber\\
&&<
\begin{cases}
\frac{4k_\gamma}{L}& (L\ll L_{\rm att,\alpha/\beta})\\
\frac{\sqrt{\sqrt{2}-1} k_\gamma}{L_{\rm att,\alpha/\beta}} & (L\gg L_{\rm att,\alpha/\beta})\\
\end{cases}\label{eq:ma_width_bragg}.
\end{eqnarray}
The right hand side of Eq.~(\ref{eq:ma_width_bragg}) for $L\sim L_{\rm att,\alpha/\beta}$ can be approximated to $\frac{4k_\gamma}{L}$ without a large loss of precision. When the Bragg condition and the resonant condition (Eq.~(\ref{eq:resonant_bragg})) are fulfilled, the conversion probability can be evaluated as 
\begin{equation}
P_{a\leftrightarrow\gamma}=\left|\frac{S^a_N}{T_0}\right|^2\simeq\frac{1}{4}(\zeta_{TS}N)^2\left(\frac{1-{\rm exp}(-{\rm Im}(\eta_0\mp\eta)N}{i {\rm Im}(\eta_0\mp\eta)N}\right)^2.
\label{eq:prob_bragg_0}
\end{equation}
It is convenient to represent the amplitudes $\eta$ and $\zeta$ by means of macroscopic parameters of the crystal. The imaginary part of the forward X-ray scattering amplitude causes X-ray absorption determined by the relation,
\begin{equation}
{\rm Im}\eta_{0} N=\frac{\mu H}{2{\rm cos}\theta_{\rm B}},
\label{eq:eta_macro}
\end{equation}
where $\mu=\frac{1}{L_{\rm att}}$ is the absorption coefficient of the crystal. 
On the other hand, the conversion amplitude $\zeta^{(\prime)}_{jk}$ is related to the effective electric fields $E_T$ in the crystal. $\zeta_{TS} N$ and $\zeta^\prime_{TS} N$ can be rewritten to be 
\begin{eqnarray}
\zeta^{(\prime)}_{TS} N&=& \frac{1}{2}g_{a\gamma\gamma}E_T HD,\label{eq:zeta_macro0}\\
D&\equiv&
\begin{cases}
\frac{k_a}{q_T}\frac{{\rm sin}(\theta^\gamma_T+\theta^a_S)}{{\rm cos}\theta^a_S} & (\gamma\ \to \ a)\\
\frac{k_a^2}{k_\gamma q_T}\frac{{\rm sin}(\theta^a_T+\theta^\gamma_S)}{{\rm cos}\theta^\gamma_S} & (a \ \to \ \gamma)
\end{cases},
\label{eq:zeta_macro}
\end{eqnarray} 
where the effective field is defined as
\begin{equation}
E_{\rm T}=q_{\rm T}\frac{F_c}{V}\int d^3{\bf r}\phi({\bf r}){\rm exp}(i{\bf q}\cdot {\bf r}).
\end{equation}
The factor, $D$, can be reduced to unity under the Bragg condition ($k_\gamma\simeq k_a$ and $\theta^{\gamma/a}_{T/S}\simeq\theta_{\rm B}$). Since the integral is roughly proportional to $q_T^{-2}$, $E_T$ is inversely proportional to $q_T$. The uncertainty due to the approximation $k_\gamma\simeq k_a$ is $\mathcal{O}(0.1\%)$ for $k_\gamma=\mathcal{O}(10\ {\rm keV})$ and $m_a=\mathcal{O}(1\ {\rm keV})$. From Eqs.~(\ref{eq:prob_bragg_0}, \ref{eq:eta_macro}, \ref{eq:zeta_macro0}, \ref{eq:zeta_macro}) and ${\rm Im}\eta=\kappa{\rm Im}\eta_0$,  
\begin{eqnarray}
P_{a\leftrightarrow \gamma}&=&\left(\frac{1}{4}g_{a\gamma\gamma}E_T L_{\alpha/\beta}{\rm cos}\theta_{\rm B}\right)^2 \nonumber \\
&=&8.6\times 10^{-8}\nonumber\\
& \times& \left(\frac{g_{a\gamma\gamma}}{10^{-3}\ {\rm GeV}^{-1}}\frac{E_T}{10^{11}\ {\rm V/m}}\frac{L_{\alpha/\beta}{\rm cos}\theta_{\rm B}}{1\ {\rm mm}} \right)^2, \label{eq:prob_bragg}\\
L_{\alpha/\beta} &\equiv& 2L _{\rm att,\alpha/\beta}\left(1-{\rm exp}\left(-\frac{L}{2L_{\rm att,\alpha/\beta}}\right)\right),
\end{eqnarray}
where $L_{\alpha/\beta}$ are effective conversion lengths of the Bloch waves. This expression is quite analogous to the result of the Bragg-case conversion as shown in Eq.~(\ref{eq:bragg_prob}). However, the penetration length $L_{\rm B}\sim\ \mathcal{O}({\rm \mu m})$ is replaced by the effective conversion length $L_{\rm \alpha/\beta}\sim L_{\rm att} \sim \mathcal{O}({\rm 100\ \mu m})$. Therefore, the probability of the Laue-case conversion is larger than that of the Bragg-case conversion. In particular, Bormann effect \cite{bib:borrmann1, bib:borrmann2} can enhance further the conversion probability from the Bloch wave $\alpha$ by $\frac{1}{1-\kappa}\sim\mathcal{O}(10)$.

Here, we consider the crystal dependence of the conversion probability. Since $E_T$ is approximately inversely proportional to $q_T$, lattice planes with lower indexes have higher $E_T$. The coefficient  $F_c$ for diamond-like crystals can be described as follows,
\begin{equation}
  F_c = \begin{cases}
    8 & (h,k,l\ {\rm are\ all\ odd/even}, h+k+l=4m) \\
    4\sqrt{2} & (h,k,l\ {\rm are\ all\ odd/even}, h+k+l=4m\pm 1)\\
    0 & ({\rm otherwise})
  \end{cases},
\end{equation}  
where $(hkl)$ is the indexes of reflecting lattice plane, and $m$ is an integer. The dependency of $F_c$ favors lattice planes with $h+k+l=4m$. Therefore, the (220) lattice plane is the most suitable one with the highest effective field. The species of crystals also strongly influence on the conversion probability through $E_T$ and $L_{\rm att}$. Table~\ref{tab:crystals} shows parameters of C (diamond), Si and Ge (220) lattice planes for X rays with a photon energy of 20 keV as an example. Diamond crystals have the most ideal specifications among them, with a long attenuation length, high $\kappa$, and high $E_T$. Although a simplified calculation without X-ray absorption \cite{bib:simpletheory} favors higher-$Z$ crystals with higher $E_T$ such as Ge and W, our realistic calculation prefers low-$Z$ crystals (C and Si) because they have longer effective conversion lengths of the Laue-case conversion.

\begin{table}[htb]
  \begin{center}
    \caption{The summary of parameters of C (diamond), Si and Ge(220) lattice planes for X rays with the photon energy of 20 keV.}
    \begin{tabular}{|c||c|c|c|c|c|c|} \hline
      crystal & $q_T$ [keV] & $\theta_{\rm B} [{\rm deg}]$ & $L_{\rm att}$ & $\kappa$ & $E_T$ [V/m] & $m_\gamma$ [eV]\\ \hline \hline
      C(220) & 9.83 & 14.2 & 13 mm & 0.981 & $6.8 \times 10^{10}$ & 38 \\ \hline
      Si(220) & 6.46 & 9.3 & 1.1 mm & 0.969 & $4.4 \times 10^{10}$ & 31 \\ \hline
      Ge(220) & 6.20 & 8.9 & 43 ${\rm \mu m}$ & 0.965 & $7.3 \times 10^{10}$ & 44 \\ \hline
    \end{tabular}
     \label{tab:crystals}
  \end{center}
\end{table}

\begin{figure}
  \centering
  \includegraphics[width=0.5\textwidth, angle=0]{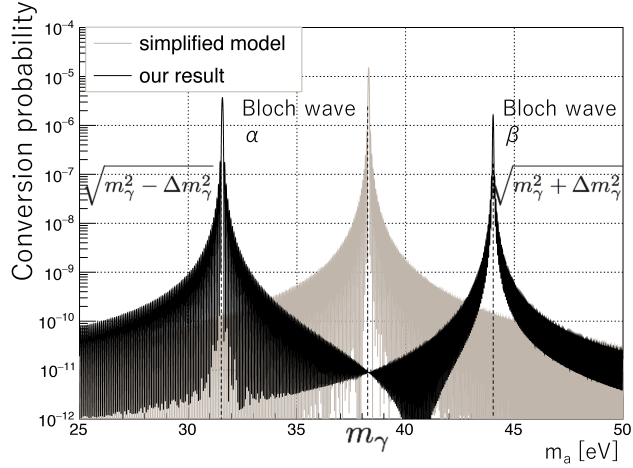}
  \caption{Conversion probabilities by a C(220) 10 mm-thick single crystal under the Bragg condition as a function of $m_a$ (the black line). Injected X rays are simple plane waves, the photon energy of X rays is 20 keV, and we assume $g_{a\gamma\gamma}=10^{-3}\ {\rm GeV}^{-1}$. The result of simplified model done in Ref.~\cite{bib:simpletheory} is also shown by the gray line.  }
   \label{fig:conv}
\end{figure}

From Eq.~(\ref{eq:xtoalps}), the $m_a$ dependence of the conversion probability is calculated numerically. The results under the Bragg condition ($\theta^\gamma_T=\theta_{\rm B}$) are shown as the black line in Fig.~\ref{fig:conv}. In this calculation, we assume that injected X rays are simple plane waves, the X-ray photon energy is 20 keV, the converter is a C(220) 10 mm-thick single crystal and the coupling constant is $g_{a\gamma\gamma}=10^{-3}\ {\rm GeV}^{-1}$. The assumed value of $g_{a\gamma\gamma}$ is the maximum sensitivity of previous LSW experiments around $m_a=100$~eV. The result of simplified model done in Ref.~\cite{bib:simpletheory} is also shown by the grey line. As shown in Eqs.~(\ref{eq:ma_bragg}, \ref{eq:ma_width_bragg}), the conversion probability has separated peaks corresponding to the Bloch waves $\alpha$ $\left (\sqrt{m_\gamma-\Delta m_\gamma^2}=32\ {\rm eV}\right)$ and $\beta$ $\left (\sqrt{m_\gamma+\Delta m_\gamma^2}=44\ {\rm eV}\right)$, and their full widths at half maximum of $(P_{\gamma\leftrightarrow a})^2$ are determined from Eq.~(\ref{eq:ma_width_bragg}) as 35 and 48 meV, respectively. The Borrmann effect makes the peak probability of the Bloch wave $\alpha$ larger than that of $\beta$. The conversion probability has an oscillation structure and sub peaks around the main peak due to the phase factor in Eq.~(\ref{eq:xtoalps}). The oscillating contribution from the Bloch waves $\alpha/\beta$ interfere destructively at the plasma frequency $m_\gamma=38\ {\rm eV}$. Although the result of simplified model also has a structure similar to our calculation, there are two major differences between them. While our result has separated peaks, the simplified calculation without X-ray diffraction has a single peak at $m_\gamma=38\ {\rm eV}$, which is the most insensitive mass in our result. Moreover, the peak probabilities of our result are smaller than that of the simplified calculation because X-rays are divided into two waves and absorbed partially by the crystal. These effects are particularly important for the conversion of ALPs with $m_a\sim m_\gamma$.

\subsubsection{Where the incident angle is away from the Bragg angle}
\begin{figure}
  \centering
  \includegraphics[width=0.5\textwidth, angle=0]{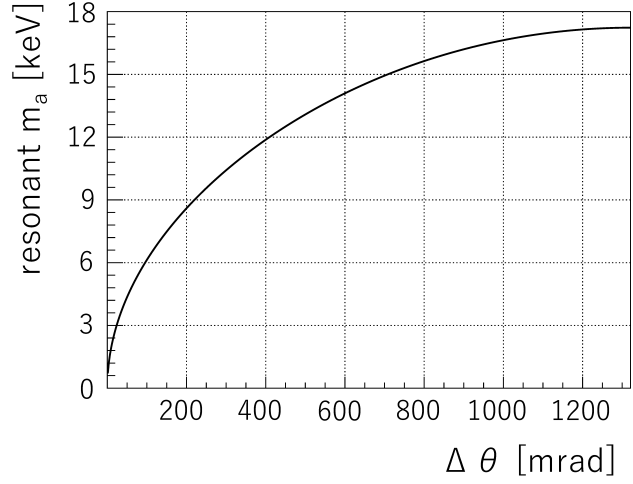}
  \caption{The most sensitive $m_a$ as a function of the detuning angle $\Delta\theta$. The converter is a C(220) crystal, and the photon energy of X rays is 20~keV. The upper limit of $\Delta\theta$=1.32~rad corresponds to $\theta^\gamma_T=\frac{\pi}{2}$.}
   \label{fig:sensitive_mass}
\end{figure}

In this subsection, we consider the case where the incident angle is much larger than $\theta_{\rm B}$ ($\theta^\gamma_T\gg \theta_{\rm B}$). This means $u\simeq\Delta\phi_\gamma$, $C_\alpha\simeq 0$ and $C_\beta\simeq 1$ in Eqs.~(\ref{eq:qx}, \ref{eq:dphi_gg}, \ref{eq:c_approx}, \ref{eq:u_approx}). In this case, the conversion probability is determined only by the contribution of the Bloch wave $\beta$, the second term of Eq.~(\ref{eq:xtoalps}). The conversion in this case is resonant under the condition, 
\begin{eqnarray}
&&\left({\rm Re}(\eta_0)-2\Delta\phi_a\right)N=0\nonumber\\
\leftrightarrow m_a&=& \sqrt{m_\gamma^2+2q_T\left(k_\gamma{\rm sin}\theta^\gamma_T-\frac{q_T}{2}\right)}\nonumber\\
&\simeq& \sqrt{m_\gamma^2+2q_Tk_\gamma{\rm cos}\theta_{\rm B}\Delta \theta}.\label{eq:ma_away}
\label{eq:width}
\end{eqnarray}
Here, we use the relation of Eq.~(\ref{eq:delta_phi_a}), and $\Delta\theta\equiv\theta^\gamma_T-\theta_{\rm B}$ is the detuning angle. The resonance has a range of $m_a$ similar to Eq.~(\ref{eq:ma_width_bragg}), 
\begin{eqnarray}
&&\left|m_a^2-m_\gamma^2-2q_T\left(k_\gamma{\rm sin}\theta^\gamma_T-\frac{q_T}{2}\right)\right|\ltsim\frac{4k_\gamma}{L}.\label{eq:ma_width_away}
\end{eqnarray}
The right hand side of Eq.~(\ref{eq:ma_width_away}) is approximated according to the discussion below Eq.~(\ref{eq:ma_width_bragg}), and the expression can be reduced to the one in Ref.~\cite{bib:simpletheory} when $L$ is much shorter than $L_{\rm att}$. From Eq.~(\ref{eq:ma_width_away}), the acceptable angular divergence of injected X rays/ALPs is calculated to be 
\begin{equation}
\Delta \theta_{\rm CV}=\frac{2d}{\pi L{\rm cos}\theta_{\rm B}},\label{eq:tcv}
\end{equation}
where the angular divergence is defined as a full width at half maximum of $(P_{\gamma\to a})^2$. As shown in Eq.~(\ref{eq:ma_away}), the resonant mass strongly depends on the deviation of the incident angle from the Bragg angle, $\Delta\theta$. The resonant mass $m_a$ has its maxima $m_a\simeq\sqrt{2q_T (k_\gamma-\frac{q_T}{2})}$ when injected X rays are nearly parallel to the crystal surface ($\theta^\gamma_T\simeq\frac{\pi}{2}$). Figure~\ref{fig:sensitive_mass} shows the sensitive $m_a$ as a function of $\Delta\theta$ in the case when the converter is a C(220) crystal, and X-ray photon energy is 20~keV. Although our results deviates from that of Ref.~\cite{bib:simpletheory} under the Bragg condition as shown in the previous subsection, the sensitive mass for $\Delta\theta\gg 0$ is almost the same. The sensitive mass reaches 1 and 10~keV when $\Delta\theta$ is 2.62 and 275~mrad, respectively. The resonant mass has its maximum value of $\sim 17.2$ keV when $\Delta\theta=1.32$ rad ($\theta^\gamma_T=\frac{\pi}{2}$).

\begin{figure}
  \centering
  \includegraphics[width=0.5\textwidth, angle=0]{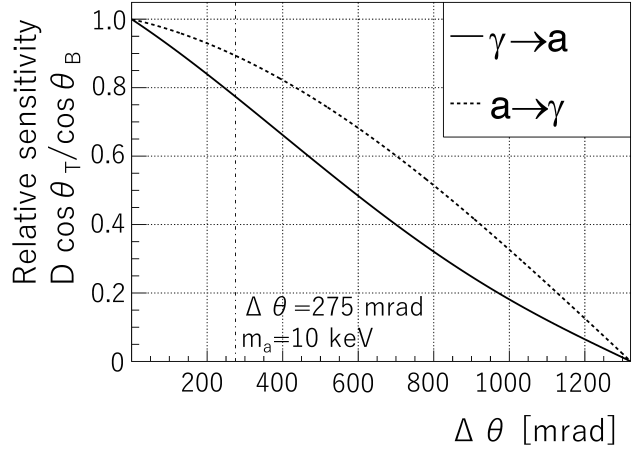}
  \caption{The $\Delta\theta$ dependence of the factor $D{\rm cos}\theta_T$. The factor corresponds to the relative sensitivity to $g_{a\gamma\gamma}$. The solid and dotted lines show the factor for the conversion and the re-conversion, respectively. The factors are normalized to their maxima, ${\rm cos}\theta_{\rm B}$ ($\Delta\theta=0$).}
   \label{fig:factorD}
\end{figure}

\begin{figure}
  \centering
  \includegraphics[width=0.5\textwidth, angle=0]{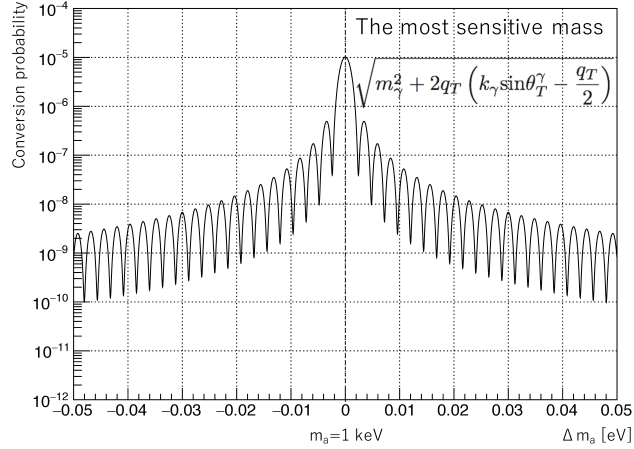}
  \caption{The conversion probability when the detuning angle is $\Delta \theta=2.62\ {\rm mrad}$. The horizontal axis shows the mass deviation from the most sensitive mass $m_a=1$ keV. Other conditions are the same as Fig.~\ref{fig:conv}}
   \label{fig:detuning_prob}
\end{figure}

Under the resonant conversion condition of Eq.~(\ref{eq:width}), the conversion probability $P_{a\leftrightarrow \gamma}$ can be calculated in the similar way as Eq.~(\ref{eq:prob_bragg}) except that the approximation of $k_\gamma\simeq k_a$ and $\ \theta^{\gamma/a}_{T/S}\simeq \theta_B$ cannot be generally justified, 
\begin{eqnarray}
P_{a\leftrightarrow \gamma}&=&\left(\frac{1}{2}g_{a\gamma\gamma}E_T L_{\rm eff}D{\rm cos}\theta_{\rm T}
\right)^2 \nonumber \\
&=&3.5\times 10^{-7}\times (D{\rm cos}\theta_T)^2\nonumber\\
&& \times \left(\frac{g_{a\gamma\gamma}}{10^{-3}\ {\rm GeV}^{-1}}\frac{E_T}{10^{11}\ {\rm V/m}}\frac{L_{\rm eff}}{1\ {\rm mm}} \right)^2,\\
L_{\rm eff} &\equiv& 2L _{\rm att}\left(1-{\rm exp}\left(-\frac{L_{T}}{2L_{\rm att}}\right)\right),
\end{eqnarray}
where $L_{\rm eff}$ is an effective conversion length of the crystal, $L_T=\frac{H}{{\rm cos}\theta_T}$ is the path length, and $\theta_T=\theta^{\gamma/a}_T$ is the angle of injected beams. The pre-factor is $2^2$ times larger than Eq.~(\ref{eq:prob_bragg}) because X rays are not divided into the Bloch waves $\alpha/\beta$ in this case. The sensitivity to the coupling constant $g_{a\gamma\gamma}$ is proportional to the factor, $D{\rm cos}\theta_T$, which depends on the detuning angle $\Delta\theta$. The $\Delta\theta$ dependence of the factor, $D{\rm cos}\theta_T$, for the conversion and the re-conversion is shown by Fig.~\ref{fig:factorD}. These factors are normalized to their maxima, ${\rm cos}\theta_{\rm B}$ ($\Delta\theta=0$). The factor decreases monotonically and vanishes at $\theta^\gamma_T=\frac{\pi}{2}$. The factor of the re-conversion is higher than that of the conversion due to ${\rm cos}\theta^\gamma_T < {\rm cos}\theta^a_T$ according to the discussion below Eq.~(\ref{eq:fresnel}). Figure~\ref{fig:detuning_prob} shows the conversion amplitude numerically calculated from Eq.~(\ref{eq:xtoalps}) as a function of $m_a$. In this calculation, we assume the same condition as Fig.~\ref{fig:conv}, except that the injection angle is detuned by $\Delta\theta=2.62$ mrad, which corresponds to the resonant mass of 1~keV. In contrast to the case under the Bragg condition, the conversion probability has a narrow peak with the mean value of $m_a=1$ keV and the width of 1.5 meV, which are determined by Eqs.~(\ref{eq:ma_away}, \ref{eq:ma_width_away}).

\section{experimental setup and sensitivity}
\begin{figure}
  \centering
  \includegraphics[width=0.5\textwidth, angle=0]{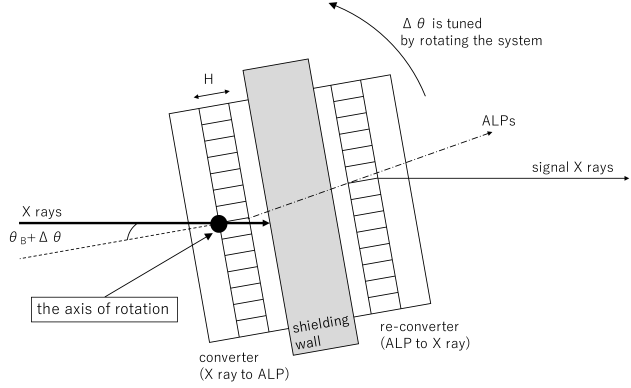}
  \caption{The schematics of an experimental setup for a LSW experiment utilizing a resonant X ray-ALP conversion by atomic electric field. The detuning angle $\Delta\theta$ is tuned by rotating the system. The axis of rotation is located at the X-ray injection point.}
   \label{fig:setup}
\end{figure}

Based on the above calculation, we design a new LSW experiment using atomic electric fields. Figure~\ref{fig:setup} shows an experimental setup using the Laue-case conversion in crystals. The converter and re-converter are two independent crystal blades, or a channel-cut crystal with two blades fabricated on a monolithic single crystal. Their reflecting lattice planes are perpendicular to the surfaces of blades. The lattice planes of the converter and re-converter is required to be parallel to each other with the precision of Eq.~(\ref{eq:tcv}). The alignment procedure can be more easily performed by using a channel-cut crystal whose parallelism between lattice planes is exact. The alignment of crystals can be guaranteed by measuring the intensity of reflected X rays as a function of $\Delta\theta$ (a rocking curve). Between two blades, a shielding wall is installed to block unconverted X rays. 

X rays are injected to the converter with an incident angle of $\theta_{\rm B}+\Delta \theta$. When ALPs' mass satisfies Eq.~(\ref{eq:width}), the Laue-case conversion from X rays to ALPs takes place. These ALPs pass through the shielding wall without absorption, and some of them are subsequently re-converted to signal X rays by the re-converter. If the parallelism between reflecting lattice planes of these crystals is guaranteed, the resonant condition of the re-conversion is automatically satisfied when incident X rays are resonantly converted into ALPs.
Finally, signal X rays are measured by an X-ray detector. As shown in the previous section, the detectable ALPs' mass $m_a$ depends on $\Delta\theta$. In contrast to other ALPs searches, the sensitive mass can be tuned quite easily by rotating crystals slightly.

From the calculations shown above, the number of detectable signal X rays may be represented as follows,
\begin{eqnarray}
N_{\rm obs}=N_\gamma T_{\rm DAQ} \frac{\Delta \theta_{\rm CV}}{\Delta \theta_{\rm BL}}\epsilon_{\rm d}P_{a\leftrightarrow\gamma}^2,
\end{eqnarray}
where $N_\gamma$ is an X-ray flux, $T_{\rm DAQ}$ is a data acquisition time, $\Delta\theta_{\rm BL}$ is the effective angular divergence of X rays including their energy bandwidth, and $\epsilon_{\rm d}$ is the detection efficiency. The sensitivity to $g_{a\gamma\gamma}$ (90\% C. L.) when backgrounds do not exist is shown by the following formula,
\begin{eqnarray}
g_{a\gamma\gamma} &>& 2 \times 10^{-4} \left(\frac{N_\gamma}{10^{13}\ {\rm Hz}}\frac{T_{\rm DAQ}}{10^3\ {\rm s}}\frac{\Delta \theta_{\rm CV}}{\Delta \theta_{\rm BL}}\epsilon_{\rm d}\right)^{-\frac{1}{4}} \nonumber\\
&&\times\left(\frac{E_T}{10^{11}\ {\rm V/m}}\frac{L_{\rm eff}}{1\ {\rm mm}}D{\rm cos}\theta_{\rm T}\right)^{-1},
\end{eqnarray}
for $m_a\simeq\sqrt{m_\gamma^2 + 2q_Tk_\gamma{\rm cos}\theta_{\rm B}\Delta\theta}$ with the bandwidth determined by $\Delta \theta_{\rm BL}$. 

The experimental setup requires an X-ray source with high photon flux. Third-generation large-scale synchrotron radiation facilities are the most suitable X-ray sources for the experiment since these facilities are optimized to generate intense X-ray beams. There are four X-ray facilities of this kind in the world today: PETRA${\rm I\hspace{-.1em}I\hspace{-.1em}I}$, SPring-8, ESRF and APS. The maximum value of their X-ray flux is $\mathcal{O}(10^{13})$ photon/s at the photon energy of 20 keV (SPring-8 BL19LXU \cite{bib:bl19}). 

When a current third generation facility, C(220) crystals with the thickness of 10 mm and X rays with the photon energy of 20 keV ($\theta_{\rm B}=14.2$ deg) are used, $N_\gamma\sim 1\times 10^{13}\ {\rm Hz}$, $E_T=6.8\times 10^{10}\ {\rm V/m}$ and $L_{\rm eff}=8.3$~mm. The effective angular divergence $\Delta\theta_{\rm BL}$ depends on X-ray optics, and has the typical value of $\sim 30\ {\rm \mu rad}$ ($\frac{\Delta\theta_{\rm CV}}{\Delta\theta_{\rm BL}}=\frac{8.3\ {\rm nrad}}{30\ {\rm \mu rad}}=2.8\times 10^{-4}$). The experimental setup has the sensitivity of $g_{a\gamma\gamma}>3\times 10^{-4}\ {\rm GeV}^{-1}$ in the case when $\epsilon_{\rm d}=0.8$, $T_{\rm DAQ} =1500\ {\rm s}$, $\Delta\theta\simeq0$ and $D{\rm cos}\theta_T\simeq{\rm cos}\theta_{\rm B}$. Although the factor $D{\rm cos}\theta_T$ is a monotonically decreasing function of $\Delta\theta$ and $m_a$, the sensitivity loss due to the factor is less than $\sim 20\%$ for ALPs with the mass up to 10 keV as shown in Fig.~\ref{fig:factorD}. The sensitivity is much higher than prior experiments in the keV region. A wide range of $m_a$ can be searched by scanning $\Delta \theta$ with a step of $\Delta\theta_{\rm BL}\sim 30\ {\rm \mu rad}$.

\section{conclusion}
ALPs are particles predicted by theories beyond the Standard Model whose existence is theoretically and astronomically motivated. In particular, ALPs in the keV range can provide possible explanation for anomalous solar activities and unidentified X-ray emission line from galaxy clusters. 
 
In this paper, we performed the first theoretical calculation of the Laue-case conversion between X rays and ALPs in crystals. The effects of X-ray diffraction and nonzero ALPs' mass are taken into account at the same time. This paper shows that the effective conversion length of the Laue-case conversion is $\mathcal{O}(10^2)$ times longer than the penetration length $L_{\rm B}$ of the Bragg-case conversion as pointed out by {\it Buchm\"{u}ller and Hoogeveen}. The conversion probability under the Bragg condition has two maxima corresponding to the Bloch waves, and the sensitivity for ALPs with the mass $m_a$ less than the plasma frequency of crystals can be enhanced further due to the Borrmann effect. 

We also discussed a LSW experiment using the conversion scheme at a current X-ray facility. Its experimental setup is composed of diamond blades (a converter and a re-converter) and a shielding wall between them. By scanning the detuning angle, ALPs with the broad sensitive $m_a$ up to $\mathcal{O}(10$ keV) can be searched.

\begin{acknowledgements}
The research is funded by the Japan Society for the Promotion of Science (Grant number 15J00509). We would like to thank Shoji Asai for useful discussions and suggestions.
\end{acknowledgements}

\bibliography{apssamp}

\end{document}